\documentclass[12pt]{article}
\pdfoutput=1
\usepackage[latin1]{inputenc}
\usepackage{amsmath,amssymb,exscale}
\usepackage{psfrag,graphicx}
\input epsf

\def\be{\begin{equation}}
\def\ee{\end{equation}}
\def\bea{\begin{eqnarray}}
\def\eea{\end{eqnarray}}
\newcommand{\eq}[1]{Eq.~(\ref{#1})}

\def\ms{\Lambda}

\def\cmij{c^{(ij)}_m}
\def\clij{c^{(ij)}_\lambda}
\def\cmtt{c^{(22)}_m}

\def\cltt{c^{(22)}_\lambda}

\def\gappeq{\mathrel{\rlap {\raise.5ex\hbox{$>$}}
{\lower.5ex\hbox{$\sim$}}}}
\def\lappeq{\mathrel{\rlap{\raise.5ex\hbox{$<$}}
{\lower.5ex\hbox{$\sim$}}}}

\def\I{\rm 1\kern-.24em l} % Yes, I know. This ain't capital.

\begin{document}
\topmargin -1.0cm
\oddsidemargin -0.5cm
\evensidemargin -0.5cm

\pagestyle{empty}
\begin{flushright}
July 2011\\
\end{flushright}
\vspace*{5mm}

\begin{center}
\vspace{1cm}
{\Large\bf A Distorted MSSM Higgs Sector
\vspace{.3cm}\\
from Low-Scale Strong Dynamics}

\vspace{1cm}
{\large 
Tony Gherghetta$^{a,b}$ and Alex Pomarol$^c$}\\
\vspace{0.5cm}
{\it \small {$^a$School of Physics, University of Melbourne, Victoria 3010,
Australia}}\\
{\it \small {$^b$Stanford Institute of Theoretical Physics, Stanford University, Stanford, CA 94305, USA}}\\ 
{\it \small{$^c$Departament de F\'isica, Universitat Aut{\`o}noma de Barcelona, 08193~Bellaterra,~Barcelona}}
\vspace{.4cm}
\end{center}

\vspace{1cm}
\begin{abstract}
We show that when supersymmetry is broken at the TeV scale by strong dynamics, 
the Higgs sector of the MSSM can be drastically modified. This arises from possible 
sizeable mixings of the Higgs with the resonances of the strong sector. In particular
the mass of the lightest Higgs boson can be significantly above the MSSM bound
($\sim 130$ GeV). Furthermore only one 
Higgs doublet is strictly necessary, because the Yukawa couplings can have a very 
different structure compared to the MSSM. Using the AdS/CFT correspondence electroweak 
precision observables can be calculated and shown to be below experimental bounds.
The most natural way to generate sparticle masses is through mixing with the composite 
states. This causes the gauginos and Higgsinos to easily obtain Dirac masses around 
200 GeV, while scalar masses can be generated either from extra $D$-terms 
or also through mixing with the strongly-coupled states. Finally one of the most interesting 
predictions of these scenarios is the sizeable decay width of the Higgs boson into a very
light gravitino $(\sim 10^{-4}$ eV) and a Higgsino.

\end{abstract}

\vfill
\begin{flushright}
NSF-KITP-11-137\\
SU-ITP-11/38
\end{flushright}
\eject
\pagestyle{empty}
\setcounter{page}{1}
\setcounter{footnote}{0}
\pagestyle{plain}

%
%============================================================
%
\section{Introduction}

One of the most remarkable predictions of the minimal supersymmetric standard model 
(MSSM) arises in the Higgs sector. Supersymmetry requires the presence of two Higgs 
doublets $H_{1,2}$ with a fully fixed Yukawa structure. Only one Higgs $H_2$ can 
couple to the up-quark sector, while the other Higgs $H_1$ couples to the down-quark 
and lepton sectors. The Higgs potential quartic couplings are also determined in terms 
of the gauge couplings.

When supersymmetry is softly broken some of these predictions are modified by quantum 
effects, but nevertheless the Higgs quartic and Yukawa couplings can be predicted 
as a function of the soft masses. This leads to some general predictions, the most important
being the presence of a light Higgs boson with mass $\lesssim 130$ GeV. In addition there
are relations between the heavy Higgs masses such as $m_{H^\pm}^2\simeq m_A^2+ m_W^2$.
These predictions occur for the most popular models of supersymmetry-breaking, 
such as in gauge-mediated supersymmetry breaking (GMSB) models.

The purpose of this paper is to show that the Higgs sector can be drastically modified if 
supersymmetry is broken dynamically by a strong sector at low energies $\sim$ TeV.
This can occur if the Higgs responsible for electroweak symmetry breaking has sizeable 
couplings to this sector, which then leads to mixings with composite states. In particular, 
we will show that the mass of the lightest Higgs boson can not only be heavier than the 
tree-level value, $m_Z$ (without large radiative corrections as in usual MSSM models),
but can also be significantly enhanced well beyond $\sim 130$ GeV. 
Furthermore the Yukawa couplings can fully arise from supersymmetry-breaking effects, 
implying that it is possible to have a supersymmetric standard model with only one Higgs doublet.

To accomplish this, there are several obvious generic problems that need to be addressed.
First, a strong sector with a partly-composite Higgs generically leads to corrections to standard model (SM) observables that have been very well measured at collider experiments. A second problem is related with the absence of large flavor changing neutral currents (FCNC). Finally, 
we must generate sizeable gaugino and squark/slepton masses. However in contrast 
to GMSB models, the problem of generating sufficiently large Higgsino masses (the $\mu$-problem)
is easily solved in our setup because the Higgsino can have sizeable mixing with the composite states.

We will make the simplifying assumption that there is a unique strong sector responsible for supersymmetry breaking. As in GMSB models, this sector will contain fields charged under the 
SM gauge group. Due to our present lack of complete understanding of strong interactions, we will 
not consider any specific model of dynamical supersymmetry breaking. Instead we will make use 
of the AdS/CFT correspondence in order to calculate observables and obtain predictions. 
Generically this correspondence allows strongly-coupled gauge theories in the large-$N$ 
and large 't Hooft coupling limit to be described by gravitational theories in five dimensions 
\cite{Maldacena:1997re}. We will also consider the possibility of having the strong sector being responsible not only for supersymmetry breaking, but also for electroweak symmetry breaking.
In this case the MSSM Higgs obtains a vacuum expectation value (VEV) from a tadpole, as in Bosonic Technicolor models \cite{Samuel:1990dq,Carone:2006wj}.

A generic prediction of our scenario is that by mixing with the strong sector the lightest Higgs boson can become much heavier than 130 GeV without causing any conflict with electroweak precision tests (EWPT). Since the scale of supersymmetry breaking is low ($\sim$ TeV), novel ways are needed to generate sparticle masses 
consistent with collider bounds. For example, the problem of having sufficiently heavy gaugino 
masses can be remedied by marrying the gaugino with composite states thereby generating a 
Dirac gaugino. Similarly the squark and slepton masses will depend on how the MSSM matter 
sector couples to the strong sector. If there is no direct coupling, one is forced to introduce an 
extra $U(1)_X$ to mediate the breaking of supersymmetry to squarks and sleptons via flavor-independent 
$D$-terms. Alternatively, the matter fields can mix with composite states and generate sizeable soft masses. 
In this case, Yukawa couplings can be generated from the strong sector without requiring a second Higgs 
doublet, but flavor symmetries are needed to avoid FCNC.

Our novel scenario leads to extraordinary MSSM physics. For example, the lightest Higgs boson
may be observed through the golden $WW/ZZ$ decay channel, and can also have sizeable decays 
to the gravitino and Higgsino whenever the $WW$ channel is kinematically closed.
Signals of compositeness in the quark sector could also be visible at the LHC, and
a TeV-scale mass $Z'$ gauge boson, associated with the new $U(1)_X$ gauge symmetry, 
could also be easily detected through its decay to leptons.

The scenarios presented here share some similarities with models in a warped extra dimension 
with TeV-scale supersymmetry breaking \cite{Gherghetta:2000kr, Hall:2003yc}. Also recently, 
detailed strong sectors with similar properties has been proposed in Ref.~\cite{Craig:2011ev} 
using Seiberg dualities. For previous general studies see Ref.~\cite{Brignole:2003cm}.
In addition, while this article was being completed, we became aware of analogous phenomenological 
studies in Ref.\cite{Ibe:2010ig, Davies:2011mp,Mukhopadhyay:2011rp,Azatov:2011ht}, that partially overlap with our present work.
We go beyond the analysis in Ref.\cite{Azatov:2011ht} by making use of the AdS/CFT correspondence
to make quantitative predictions for large $N$ gauge theories.

\section{The strong sector with dynamical supersymmetry breaking at the TeV scale}
\label{sec1}
The strong sector will be broadly characterized by two parameters, the scale $\ms\sim$ TeV associated 
with the mass gap (or, equivalently, the mass of the first resonance), and the number of ``colors" $N$.
We will assume that supersymmetry is completely broken in the strong sector with
supersymmetry-breaking mass-splittings of order $\ms$.
The vacuum energy is therefore estimated to be
$V\sim {N\ms^4}/{(16\pi^2)}$ (assuming strong sector fields transform in the fundamental representation of the gauge group) which implies that $F$-terms are generically of order
\be
F\sim \frac{\sqrt{N}}{4\pi}\ms^2,
\label{fterm}
\ee
and similarly for $D$-terms. In certain cases it will be useful to parametrize, as usual, the breaking of supersymmetry by a dimensionless spurion superfield
\be
\eta\equiv\theta^2\frac{F}{\ms^2}\, .
\label{spurion}
\ee

The SM fields and superpartners, denoted generically as $\Phi_i$, will be assumed to couple to the strong sector linearly:
\be
{\cal L}_{\rm int}={\hat g}_i\Phi_i\, {\cal O}_{\Phi_i}\, ,
\label{mixing}
\ee
where ${\hat g}_i$ is the coupling to an operator ${\cal O}_{\Phi_i}$ of the strong sector. In particular, for the MSSM vector supermultiplet $V_i$ an interaction of the type \eq{mixing} can arise, like in GMSB models, by assuming that the strong sector contains fields charged under the SM gauge groups. In this case we have the gauge interaction
$\int d^4 \theta \, {\hat g}_iV_i {\cal J}_i$, where the superfield ${\cal J}_i$ contains the conserved current operator of the strong sector (${\cal J}_i=\bar \theta\sigma_\mu\theta J_i^\mu+\cdots$) and ${\hat g}_i$ is the gauge coupling of the different SM gauge groups. For the MSSM chiral multiplets the interaction \eq{mixing} can arise from a superpotential term
$\int d^2\theta \, {\hat g}_i\Phi_i\, {\cal O}_{\Phi_i}$, where the same notation will be used for general superfields and component fields. Notice that, apart from couplings like \eq{mixing}, supersymmetry and gauge invariance allows for extra (non-linear) interactions between the MSSM and the strong sector. 

If the strong sector is approximately conformal at energies above $\ms$, the operators 
${\cal O}_{\Phi_i}$ can be organized according to their conformal dimension
${\rm Dim}[{\cal O}_{\Phi_i}]$. The couplings ${\hat g}_i$ will then have a dimension 
$-\gamma_i$ where $\gamma_i= {\rm Dim}[{\cal O}_{\Phi_i}] +{\rm Dim}[\Phi_i]-4$.
The renormalization group (RG) equation of the dimensionless coupling, $g_i \equiv {\hat g}_i \mu^{\gamma_i}$ is then given by
\be
\mu \frac{dg_i}{d\mu}=\gamma_ig_i+\kappa_i\frac{N}{16\pi^2}g^3_i+\cdots\, ,
\label{rgeqn}
\ee
where the second term in (\ref{rgeqn}) originates from the wave-function renormalization of $\Phi_i$ in the large-$N$ limit, with $\kappa_i$ a coefficient of order one satisfying $\kappa_i>0$. We will require the value of the coupling at the mass gap scale, $g_i(\Lambda)$, where it is useful to define the quantity 
\be
\epsilon_i\sim\frac{g_i(\ms)}{{4\pi}/{\sqrt{N}}}\, ,
\label{epsilon}
\ee
up to factors of order one.
This ratio gives the coupling strength of the MSSM fields to the strong sector, normalized with respect to the strong sector couplings $\sim 4\pi/\sqrt{N}$ \cite{'tHooft:1973jz}. Equivalently, 
$\epsilon_i$ parametrizes the mixing between the MSSM fields and the composite states (resonances) associated with the operators ${\cal O}_{\Phi_i}$. For $\epsilon_i\sim 1$ the 
mixing is maximal and the MSSM particle behaves as a resonance of the strong sector.

We are interested in the case in which $-1< \gamma_i\leq 0$ such that the coupling $g_i$ is relevant ($\gamma_i<0$) or marginal ($\gamma_i=0$). 
For $-1<\gamma_i<0$, we see from \eq{rgeqn} that $g_i$ grows towards low energies, 
before reaching the fixed-point value
\be
g_i(\ms)=\frac{4\pi}{\sqrt{N}}\sqrt{\frac{-\gamma_i}{\kappa_i}}\, .
\label{irrcoup}
\ee
In this case we obtain
\be
\epsilon_i \sim \sqrt{-\gamma_i}\, .
\label{releg}
\ee
When $\gamma_i=0$, as for the MSSM gauge couplings, the solution of \eq{rgeqn} gives
\be
\frac{1}{g^2_i(\ms)}=\frac{1}{g^2_i(\Lambda_{\rm UV})}+\kappa_i\frac{N}{16\pi^2}\ln\frac{\Lambda_{\rm UV}}{\ms}\, .
\label{rung}
\ee
For ${g^2_i(\Lambda_{\rm UV})N/(16\pi^2)}\gg 1$, we have $g_i^2(\ms) \simeq \frac{16\pi^2 }{\kappa_iN \ln({\Lambda_{\rm UV}}/{\ms})}$ and then $\kappa_i/\kappa_j\simeq g^2_j(\ms)/g_i^2(\ms)$. 
In particular, for the gauge couplings we have $\kappa_2/\kappa_1\simeq \tan^2\theta_W$
that will be useful later. Also we set $\kappa_2=1$ by a redefinition of $N$.

\subsection{Electroweak symmetry breaking by an elementary Higgs}

As mentioned in the Introduction the Higgs sector can either contain one or two Higgs doublets.
Let us begin by considering the case of two Higgs doublets with opposite hypercharge, 
$H_1$ and $H_2$, and study their potential. The potential consists of a supersymmetric 
part containing not only the usual $D$-term, but also possible $F$-terms arising from 
$\int d^2\theta \,\mu H_1H_2$.
In addition there are supersymmetry-breaking 
masses arising from $\int d^4\theta \,\eta\eta^\dagger H_iH_j^\dagger$ $(i,j =1,2)$,
and quartic couplings from $\int d^4\theta \,\eta\eta^\dagger (H_iH_j^\dagger)^2$ with $\eta$ defined in \eq{spurion}.
For the neutral $CP$-even components, $h_1$ and $h_2$, we then have
\begin{equation} 
V(h_1,h_2)=\frac{1}{2}m_{11}^2 h_1^{2}+\frac{1}{2}m_{22}^2 h_2^2-m_{12}^2h_1 h_2+
\frac{1}{4}{\lambda_{11}}h_1^4+\frac{1}{4}{\lambda_{22}}h_2^4-\frac{1}{2} {\lambda_{12}}h_1^2h_2^2\, ,
\end{equation}
where
\begin{equation} 
m_{ij}^2= |\mu|^2+\cmij \epsilon_{H_i}\epsilon_{H_j}\ms^2\ , \ \ \ \lambda_{ij}=\frac{m_Z^2}{2v^2}+\clij \epsilon_{H_i}^2\epsilon_{H_j}^2 \frac{16\pi^2}{N}\, ,
\label{parest}
\end{equation}
with $m_Z$ the $Z$-boson mass, $v\simeq 246$ GeV, and the coefficients $\cmij$ and 
$\clij$ are proportional to the degree of supersymmetry breaking in the strong sector which we assume to be of order one. The parametric dependence of the terms in Eq.~(\ref{parest}) can be understood as follows. At large 
$N$ the strong sector can be described by a theory of resonances with masses $\sim \ms$ and quartic couplings $\sim16\pi^2/N$. Assuming a mixing between these resonances and 
the Higgs of order $\epsilon_{H_i}\lesssim 1$, we obtain a Higgs mass-squared of order 
$\epsilon_{H_i}\epsilon_{H_j}\ms^2$ and quartic couplings for the Higgs of order 
$\sim \epsilon^2_{H_i}\epsilon^2_{H_j} 16\pi^2/N$. This is indeed supported by the AdS/CFT 
correspondence (see Appendix), where we find, for example, that the ratio between the Higgs quartic 
coupling and the first scalar resonance coupling is given by $\simeq 0.92 \,\epsilon_H^4$.
Supersymmetric contributions from $|\mu|^2$, if induced from the strong sector, are of order 
$\sim \epsilon_{H_1}^2\epsilon_{H_2}^2\ms^2$. Compared to the supersymmetry breaking
contributions they can be neglected, and we will assume for simplicity $\mu=0$.

\begin{figure}[top]
\centering
\hspace{-0.5cm}
\includegraphics[scale=1]{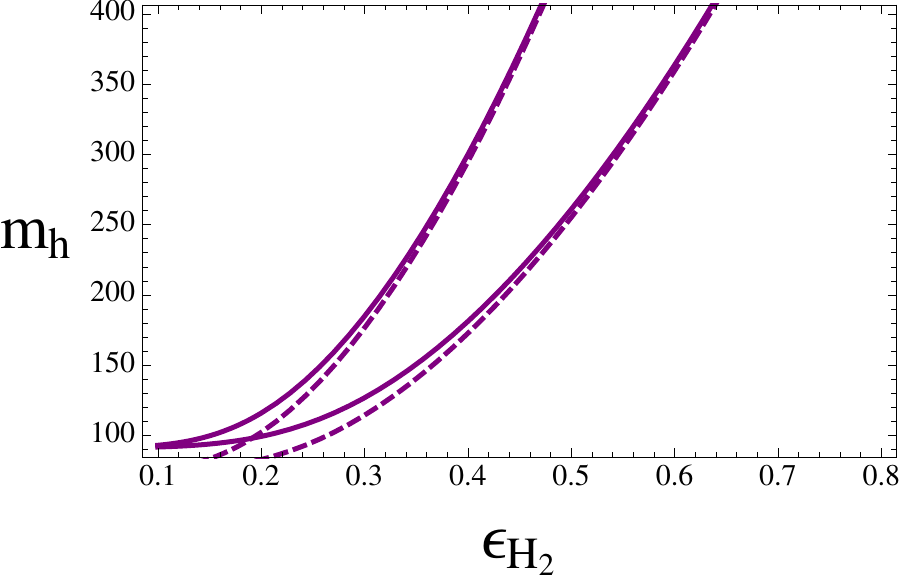} 
\caption{\footnotesize The mass (in GeV) of the lightest Higgs boson as a function of 
$\epsilon_{H_2}$ for $\cltt=1$ (upper lines) and $\cltt= 0.3$ (lower lines). 
We have taken $N=6$ and the solid (dashed) line is for tan$\beta\gg 1$ (tan$\beta= 3$).}
\label{mhplot}
\end{figure}

We will be interested in the large $\tan\beta\equiv \langle h_2\rangle/\langle h_1\rangle$ limit 
because this gives the largest possible mass for the lightest MSSM Higgs boson. This limit can be naturally achieved by taking $\epsilon_{H_1}\gg \epsilon_{H_2}$. Indeed, in this case 
$H_1$ couples stronger than $H_2$ to the supersymmetry-breaking sector, leading to the hierarchy, $m_{22}^2\ll m_{12}^2\ll m^2_{11}$, and $\tan\beta\sim \epsilon_{H_1}/\epsilon_{H_2}\gg 1$. In this limit we can integrate $h_1$ out, and obtain the effective potential for $h_2$
\begin{equation} 
V(h_2)=\frac{1}{2}m_{22}^2 h_2^2+\frac{1}{4}{\lambda_{22}}h_2^4\, ,
\label{pot2}
\end{equation}
where the coefficients $\cmtt$ and $\cltt$ have been redefined to absorb order-one corrections. 
The potential \eq{pot2} is obviously equivalent to the potential of a one-Higgs doublet model with $h_2$ playing the role of the SM Higgs. For $m_{22}^2<0$ we have electroweak symmetry breaking with $\langle h_2\rangle\simeq v$ and the Higgs mass given by
\begin{equation} 
m_{h}=\sqrt{2 \lambda_{22}}\, v\, .
\label{mhiggs}
\end{equation}
To have a sizeable effect on the Higgs mass we must consider values of $\epsilon_{H_2}>0.1$ such that corrections to $\lambda_{22}\propto \epsilon_{H_2}^4$ become sizeable.
This is shown in Fig.~\ref{mhplot} where we plot the tree-level Higgs mass for $N=6$ and 
$\cltt=1$ and $0.3$. In particular, we find that for $\epsilon_{H_2}\sim 0.3$ the tree-level Higgs mass 
can be as large as 190 GeV. 
For smaller values of $\tan\beta$ the Higgs mass, 
as expected, decreases as depicted in Fig.~\ref{mhplot}.

However, the stronger the Higgs couples to the supersymmetry-breaking sector, the larger 
the corrections to the electroweak observables. These corrections arise from the strong sector that, coupled to the Higgs, can generate large effects on the SM gauge boson self-energies.
The main contributions are parametrized by the $S$ and $T$ parameters defined in 
Ref.~\cite{Peskin:1990zt}. We will follow the notation in Ref.~\cite{Barbieri:2004qk} where 
these parameters are denoted by $\widehat S$ and $\widehat T$. These contributions can
be parametrically written as
\begin{eqnarray} 
\label{stest0}
\widehat T&=&c_T\, \frac{16\pi^2v^2\epsilon_{H_2}^4}{N\ms^2}\, , \\
\widehat S&=&c_S \, \frac{ m^2_W}{\ms^2}\epsilon_{H_2}^2\, ,
\label{stest}
\end{eqnarray}
where $c_{T,S}$ are parameters of order one.
In order to make precise predictions we will take the values of $c_{T,S}$ as
predicted by the AdS/CFT correspondence.
In the regime $\gamma_{H_2}\leq 0$, we have \cite{Round:2010kj}:
\begin{equation} 
c_T\simeq \frac{9\pi^2}{64\kappa_1(\epsilon_{H_2}^2 + 1) (2 \epsilon_{H_2}^2 + 1)}\ ,\ \ \ 
c_S \simeq\frac{9\pi^2((\epsilon_{H_2}^2 + 1)^2 - 1)}{32(\epsilon_{H_2}^2 + 1)^2\epsilon_{H_2}^2}\, ,
\label{csct}
\end{equation}
where we identify $\epsilon_{H_2}\equiv\sqrt{-\gamma_{H_2}}$\
\footnote{For $\gamma_{H_2}\rightarrow 0$ this formula changes to $\epsilon_{H_2}^2\rightarrow 1/\ln(\Lambda_{\rm UV}/\ms)$.}.
We will take $\kappa_1/\kappa_2\simeq 1/\tan^2\theta_W$ and $\kappa_2=1$ 
as mentioned earlier. 

\begin{figure}[top]
\centering
\includegraphics[scale=1.1]{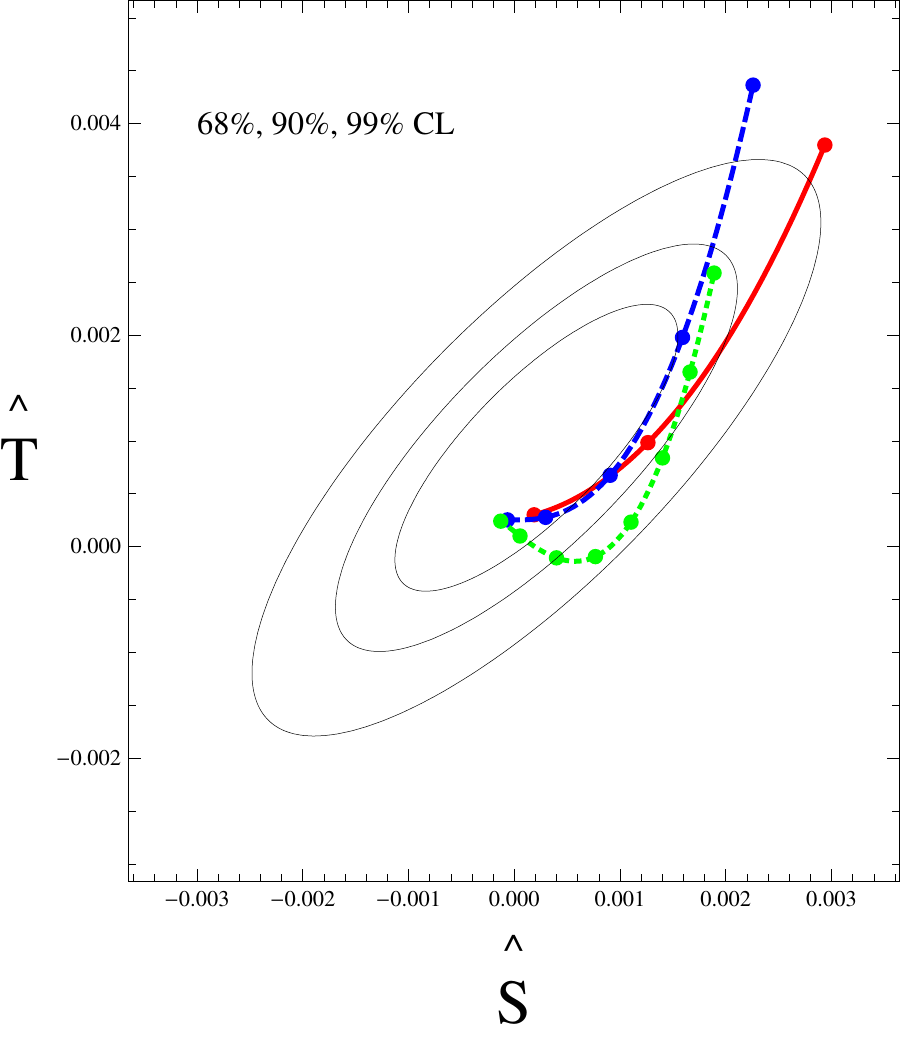} 
\caption{\footnotesize The contribution to $\widehat S$ and $\widehat T$ for $N=6$ and different values of the degree of compositeness $\epsilon_{H_2}$. The solid red line corresponds to $\ms=1$ TeV, the dashed blue line is for $\ms=2$ TeV, and the dotted green line is for $\ms=4$ TeV. The values of $\epsilon_{H_2}$ increase as we 
move out from the ellipses along the lines. The first dot (inside the inner ellipse) corresponds to $\epsilon_{H_2}=0.1$, and each successive dot represents an 
increase of $0.1$.}
\label{STch}
\end{figure}

The contributions to $\widehat S$ and $\widehat T$ are shown in Fig.~\ref{STch} for $N=6$.
The solid red line corresponds to $\ms=1$~TeV for different values of $\epsilon_{H_2}$.
The origin of the $\widehat S$-$\widehat T$ plane corresponds to the SM with a reference
Higgs mass of $m_{h, ref}=120$ GeV. Since our SM-like Higgs $h_2$ has a mass different 
from $m_{h, ref}$, we have included this effect in $\widehat S$ and $\widehat T$. In particular, 
the mass of $h_2$ is computed, as a function of $\epsilon_{H_2}$, from \eq{mhiggs} with $\cltt=1$.
In Fig.~\ref{STch} we see that, as expected, increasing $\epsilon_{H_2}$ couples 
$H_2$ stronger to the supersymmetry-breaking sector, causing the contribution to $\widehat S$ 
and $\widehat T$ to increase. Nevertheless, it is interesting to note that both contributions are positive, allowing them to remain inside the 99$\%$ CL ellipse for values as large as $\epsilon_{H_2}\sim 0.3$. For these values of $\epsilon_{H_2}$ the Higgs mass (Fig.~\ref{mhplot}) receives important contributions.

Therefore a Higgs mass well above the tree-level MSSM value, $m_Z$ is possible without affecting the EWPT of the SM\,\footnote{Of course, 
the one-loop MSSM corrections must also be added to these tree-level values,
further increasing the Higgs mass.}. As we increase $\ms$, the contributions to $\widehat S$ and $\widehat T$ becomes smaller, as is clear from the expressions (\ref{stest0}) and (\ref{stest}).
This can allow for a Higgs with a larger value of $\epsilon_{H_2}$ and therefore a larger 
mass. For $\ms=2$ TeV ($4$ TeV) the dashed blue (dotted green) line in Fig.~\ref{STch}
shows the contribution to $\widehat S$ and $\widehat T$ for several values of 
$\epsilon_{H_2}$. For $\epsilon_{H_2}\simeq 0.5$ the Higgs mass can obtain a very large value $m_{h}\simeq 450$ GeV. 

\begin{figure}
\centering
\hspace{-0.5cm}
\includegraphics[scale=1]{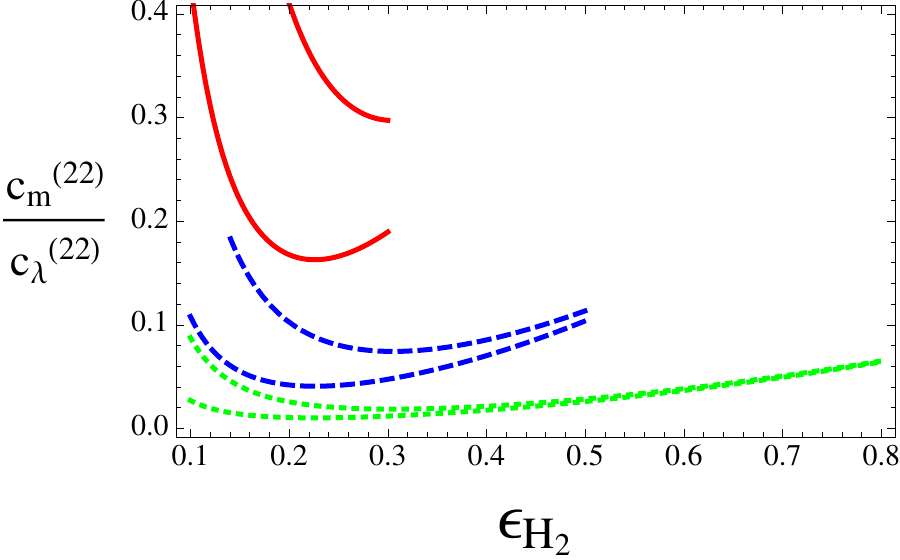} 
\caption{\footnotesize 
The degree of tuning, as defined in \eq{tun}, as a function of $\epsilon_{H_2}$.
The solid red line is for $\ms=1$ TeV, the dashed blue line is for $\ms=2$ TeV, 
and the dotted green line is for $\ms=4$ TeV.
We have taken $N=6$ and $\cltt=1$ (lower lines) 
and $\cltt=0.3$ (upper lines). }
\label{tplot}
\end{figure}

Although increasing the values of $\ms$ and/or $\epsilon_{H_2}$ can lead, as we have 
seen, to large values of the Higgs mass while still maintaining consistency with EWPT,
it is clear that a large $\ms$ and $\epsilon_{H_2}$ leads to a large $m_{22}$,
unless we tune $\cmtt\ll 1$. This tuning, which is reminiscent of the hierarchy problem, 
can only be technically natural if also $ \cltt \sim \cmtt \ll 1$, since this would correspond 
to the supersymmetric limit $F\ll \ms^2$. Therefore, for a given value of $\ms$ and 
$\epsilon_{H_2}$ we can estimate the fine tuning of the model by calculating $\cmtt/\cltt$.
Using $v^2=m_{22}^2/\lambda_{22}$ this ratio can be written as
\be
\frac{\cmtt}{\cltt}=\frac{ m_h^2}{2\cltt\epsilon_{H_2}^2\ms^2}\, .
\label{tun}
\ee
In Fig.~\ref{tplot} we plot \eq{tun} as a function of $\epsilon_{H_2}$. For $\ms=1$ TeV 
we see that no tuning is really needed, while for $\ms=2$ TeV the tuning is around 10$\%$.
Notice that the tuning is very sensitive to the coefficient $\cltt$ and is reduced when this coefficient is smaller than one, as can be appreciated in Fig.~\ref{tplot}.
Of course, smaller values of $\cltt$ lead to smaller Higgs masses as 
shown in Fig.~\ref{mhplot}. It is also interesting to remark that the tuning does not increase as the Higgs becomes more strongly coupled to the supersymmetry-breaking sector. We can then 
envisage a scenario with $\ms\sim$ 4 TeV and $\epsilon_{H_2}\sim 0.8$, certainly with some 
degree of tuning (but not higher compared to other MSSM scenarios such as GMSB), 
in which the Higgs is quite heavy, $m_h\sim$ TeV, and therefore difficult to be seen at the LHC.

\subsubsection{An alternative viewpoint}
\label{alvi}
The elementary Higgs scenario presented in the previous section can also be described using an alternative (dual) description. The fields $H_{1,2}$ can be thought of as arising from the strong sector with the identification $\epsilon_{H_{1,2}}^2={\rm Dim}[H_{1,2}]-1$. Taking $\epsilon_{H_i}^2\ll 1$ then corresponds to the limit in which these operators have dimension close to one. In this limit $H_{1,2}$ become free fields and decouple from the strong sector. This is an equivalent description of the elementary Higgs scenario and gives rise to the same physics. From this perspective there is no elementary Higgs and electroweak symmetry breaking would be triggered by the strong sector, just like in conformal technicolor theories \cite{Luty:2004ye}. These two descriptions have been identified in Ref.~\cite{Contino:2004vy} using the AdS/CFT correspondence.

\subsection{Electroweak symmetry breaking triggered by the strong sector}

Let us now assume that the strong sector not only breaks supersymmetry but also electroweak symmetry \cite{Samuel:1990dq,Carone:2006wj} by a condensate parametrized by a doublet with hypercharge $Y=-1$:
\begin{equation}
\Sigma=\frac{1}{\sqrt{2}}f_\Sigma e^{i\sigma_i\frac{\pi_i}{f_\Sigma} }\left(\begin{array}{c}1\\0\end{array}\right)\, ,
\end{equation}
where $\sigma_i$ are Pauli matrices.
The VEV of $\Sigma$ can be thought of as arising from a generic potential induced by the strong dynamics, like in QCD or technicolor models. The natural scale for $f_\Sigma$ is $\sim\sqrt{N}\ms/(4\pi)$.

The $\Sigma$ scalar could be identified with the $H_1$ field of the MSSM.
In this case its fermionic partner, a composite state, is responsible for canceling the anomalies
of the $H_2$ Higgsino. This case then corresponds to a particular case of the previous model
with $\epsilon_{H_1}\sim 1$, implying that $H_1$ behaves as a composite field of the strong sector. Contrary to the previous section, however, we must emphasize that here we are assuming that the VEV of $H_1$ is generated independently from the VEV of $H_2$.

The Higgs potential for $H_2$ is given by
\begin{equation} 
V(H_2)=m_{22}^2 |H_2|^2+{\lambda_{22}}|H_2|^4-(m_{2\Sigma}^2 H_2 \Sigma+h.c.)+\cdots\, ,
\label{pot2s}
\end{equation}
where $m_{22}$ and ${\lambda_{22}}$ are given by \eq{parest} (with $\mu=0$) and we have neglected terms with higher
powers of $\Sigma$ since they are irrelevant in the analysis. The novelty of this scenario is that 
$H_2$ gets a VEV induced by the tadpole arising from the last term of \eq{pot2s} that 
is of order $\sim\epsilon_{H_2} \Lambda^2 f_\Sigma$.
The electroweak symmetry is then broken by the strong sector and the elementary Higgs $H_2$ VEV, $v_2$,
giving rise to 
\begin{equation}
v=\sqrt{v_2^2+f_\Sigma^2}\, ,
\label{mhiggstc}
\end{equation}
where $\tan\beta$ can be identified with $v_2/f_\Sigma$. In this case the Higgs mass is
given by
\begin{equation} 
m_h^2=3 \lambda_{22} v_2^2+m_{22}^2\, .
\label{mh2}
\end{equation}
When the second term in \eq{mh2} is positive, it gives an additional contribution to the Higgs mass compared to \eq{mhiggs}, causing the Higgs to be heavier than in the ordinary MSSM.
Since in this case large values of $\epsilon_{H_2}$ are no longer needed to obtain 
sizeable contributions to the Higgs mass, we will assume $\epsilon_{H_2}\sim 0.1$.

The main contributions to $\widehat T$ and $\widehat S$ are no longer arising from the VEV 
of $H_2$ but instead from electroweak symmetry breaking in the strong sector, 
$\Sigma$. These contributions are therefore proportional to $f_\Sigma$ and given by
\begin{eqnarray} 
\label{stTCest1}
\widehat T&=&c_T \frac{16\pi^2 f_\Sigma^4}{N\ms^2v^2}\, , \\
\widehat S&=&c_S \frac{ m^2_W}{\ms^2}\frac{f_\Sigma^2}{v^2}\, ,
\label{stTCest2}
\end{eqnarray}
where $c_{T,S}$ are parameters of order one. The predictions from AdS/CFT are given in 
Eq.~(\ref{csct}) with the replacement $\epsilon_{H_2}^2\rightarrow {\rm Dim}[{\cal O}_\Sigma]-1$, where ${\cal O}_\Sigma$ is the operator of the strong sector that breaks electroweak symmetry. Assuming Dim$[{\cal O}_\Sigma ]=2$ and $\kappa_1\simeq 1/\tan^2\theta_W$, we have
\begin{equation} 
c_T\simeq \frac{c_S}{9}\tan^2\theta_W \simeq \frac{3\pi^2}{128}\tan^2\theta_W\, .
\label{csct2}
\end{equation}
If $f_\Sigma\sim v$, the contributions (\ref{stTCest1}) and (\ref{stTCest2}) to $\widehat T$ and 
$\widehat S$ are the same as those of technicolor models without a custodial symmetry, which 
are clearly ruled out mostly due to the $\widehat T$ parameter. This means that we either 
need $f_\Sigma\ll v$ or $f_\Sigma\ll \sqrt{N}\ms/(4\pi)$. The first option can be naturally implemented since parametrically, neglecting the $D$-term, we have $ v_2\sim f_\Sigma/\epsilon_{H_2}\gg f_\Sigma$. However, this implies that $\ms\ll$ TeV which becomes too 
small to generate proper masses for the superpartners~\footnote{Of course this is not a problem if supersymmetry-breaking masses for the MSSM superpartners arise from another sector.}.
Therefore we will need to also rely 
on having $f_\Sigma\ll \sqrt{N}\ms/(4\pi)$; this will not represent a severe fine tuning of 
the parameters of the model since supersymmetry could mildly protect $f_\Sigma$ to make 
it slightly smaller than its NDA value.

\begin{figure}
\centering
\begin{tabular}{cc}
\hspace{-0.1cm}
\includegraphics[scale=0.9]{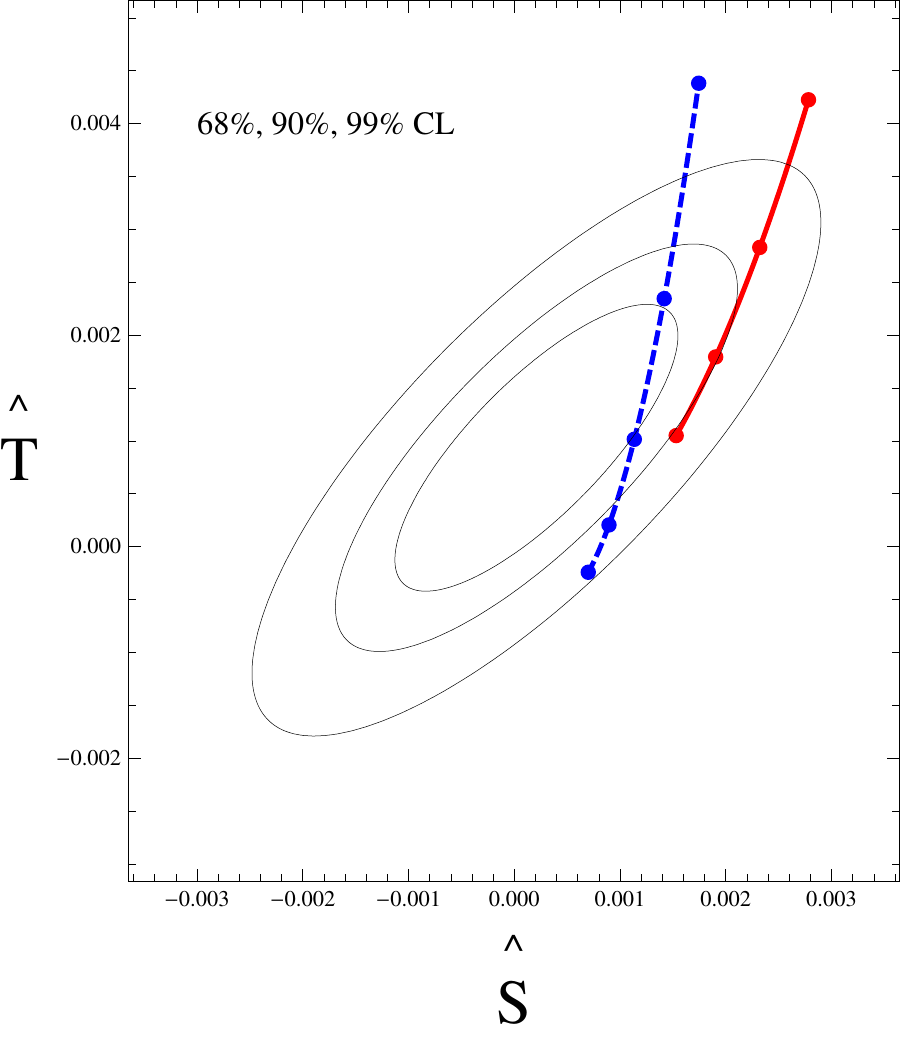} &
\end{tabular}
\caption{\footnotesize The contribution to $\widehat S$ and $\widehat T$ for $N = 6$ and different values of $f_\Sigma$. The solid red line is for $\ms=1$ TeV and the corresponding dots are for $f_\Sigma=(80,90,100,110)$ GeV. The dashed blue line is for $\ms=2$ TeV and 
the corresponding dots are for $f_\Sigma=(80,100,120,140,160)$ GeV.}
\label{STtc}
\end{figure}

In Fig.~\ref{STtc} we show the contributions to $\widehat T$ and $\widehat S$ using 
\eq{csct2}, $N=6$, and treat $f_\Sigma$ as a free parameter.  As in Fig.~\ref{STch}, the 
one-loop effects on $\widehat T$ and $\widehat S$ from a Higgs 
$h_2$ heavier than $m_{h, ref}=120$ GeV have been included. 
Its mass is calculated using \eq{mh2} with $\epsilon_{H_2}=0.1$ and $\cltt=\cmtt=1$. 
The solid red line is for $\ms=1$ TeV and the corresponding dots
are for $f_\Sigma=(80,90,100,110)$ GeV, while 
the dashed blue line is for $\ms=2$ TeV and the corresponding dots
are for $f_\Sigma=(80,100,120,140,160)$ GeV.
We can see that the model is consistent with the $\widehat S-\widehat T$ constraints
for values of $f_\Sigma$ only slightly smaller than its NDA value
$\sim \sqrt{N}\ms/{4\pi}$. 

\begin{figure}
\centering
\begin{tabular}{cc}
\hspace{-0.1cm}
\includegraphics[scale=0.9]{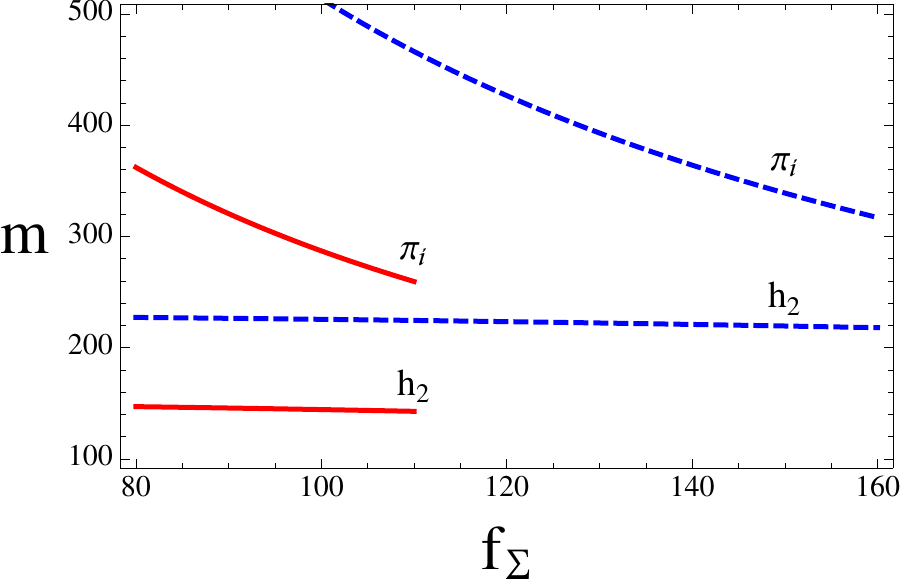} &
\end{tabular}
\caption{\footnotesize The masses (in GeV) of the lightest Higgs $h_2$ and $\pi_i$ scalars
for $N=6$, $\epsilon_{H_2}=0.1$ and $\cltt=\cmtt=1$. The solid red (dashed blue) line is for $\ms=1$ TeV 
(2 TeV).}
\label{mh2Hybrid}
\end{figure}

In these scenarios the mass of the lightest Higgs $h_2$ is dominated, for $\ms\gtrsim$ TeV, 
by the second term of \eq{mh2}, that gives 
\be
m_{h}\simeq m_{22}\sim \epsilon_{H_2}\ms\sim 200\ {\rm GeV} 
\left(\frac{\epsilon_{H_2}}{0.1}\right)\left(\frac{\ms}{2\ {\rm TeV}}\right)\, . 
\label{masshtc}
\ee
The three scalar states in the $\Sigma$ field, $\pi_i$, that corresponds to a neutral and a charged state, obtain a common mass given by
\be
m_\pi^2=\frac{m^2_{2\Sigma}v^2}{ f_\Sigma v_2}\simeq \frac{m_h^2 v^2}{f_\Sigma^2}\, ,
\ee
where $m_{2\Sigma}^2$ has been determined as a function of the other
parameters by the minimization condition. We plot in Fig.~\ref{mh2Hybrid} the mass of $h_2$ and $\pi_i$ in the regions of $f_\Sigma$ allowed by EWPT, assuming $\epsilon_{H_2}=0.1$ and $\cltt=\cmtt=1$.

The generalization of \eq{pot2s} for two elementary Higgs is straightforward. The potential
becomes
\begin{equation} 
V(H_1,H_2)=m_{ii}^2 H_iH_i^\dagger-(m_{12}^2 H_1H_2+m_{1\Sigma}^2 H_1^\dagger 
\Sigma+m_{2\Sigma}^2 H_2 \Sigma+h.c.)+ V_D+
\cdots\, ,
\label{pot12s}
\end{equation}
where $V_D$ is the ordinary MSSM $D$-term contribution.
Again, we have neglected terms that are smaller in the limit $\epsilon_{H_i}\ll 1$ and $f_\Sigma\ll v$.
We will discuss the implications of this potential below.

\section{Fermionic superpartners}

If the strong sector discussed above is the only source of supersymmetry breaking, it will
have to generate masses for all the MSSM superpartners. However, since we are assuming 
a low supersymmetry-breaking scale $\sim$ TeV, generating sufficiently heavy superpartner masses represents a challenge.

Let us begin by considering the gaugino sector. As in GMSB models, Majorana gaugino masses can be generated if the $R$-symmetry is broken. We obtain $m_{\lambda_i}\sim{g^2_i N\ms}/{(16\pi^2)} $ where $g_i$ are the gauge couplings. However for $\ms\sim$ TeV and $N\lesssim 10$ these masses seem to be too small to satisfy collider constraints, at least for gluino masses. Therefore we are forced to assume that the gauginos obtain Dirac masses. Recently GMSB models with Dirac gauginos have been studied in \cite{Benakli:2011vb}. In our setup Dirac masses can be generated by mixing the MSSM gauginos with composite fermions arising from the strong sector, just like in 
dual five-dimensional (5D) models~\cite{Gherghetta:2000kr}. In this case, the gaugino mass becomes
\begin{equation}
m_{\lambda_i}=c_{m_{\lambda_i}}g_i\frac{\sqrt{N}}{4\pi} \ms\simeq 200\ {\rm GeV}\ c_{m_{\lambda_i}}g_i \sqrt{\frac{N}{6}}
\left(\frac{ \ms}{1\ {\rm TeV}}\right)\, ,
\label{mdirac}
\end{equation}
where $c_{m_{\lambda_i}}$ are coefficients of order one.
In AdS$_5$ models (see \eq{gauginomass} in the Appendix) we find that $c_{m_{\lambda_i}}\simeq 0.6\sqrt{\kappa_i}$ . Since composite fermions marry the MSSM gauginos to 
obtain Dirac masses, we must avoid the presence of Majorana masses for these composite fermions (or at least these must be smaller than $m_{\lambda_i}$ of \eq{mdirac}).
This can be achieved by demanding that the strong sector be approximately invariant under some global symmetry
under which these composite fermions are charged. 
This could be an $R$-symmetry 
under which the spurion \eq{spurion} has charge $R=2$ and the composite fermions (in the adjoint representation of the SM gauge group) have opposite $R$-charges compared to the MSSM gauginos, $R=-1$.
An interesting possibility is to have only Dirac masses for the gluino and the wino, but not for the bino.
The bino mass will then arise from $R$-symmetry breaking terms and could be much smaller.
In the case that the $R$-symmetry is only broken by gravitational interactions, 
the bino can be lighter than the gravitino, a possibility that can be viable phenomenologically as soon as the mixing between the bino and Higgsino is smaller than $\sim 0.1$~\cite{Davies:2011mp}.

Similarly, Higgsinos can also obtain Dirac masses by mixing with composite states.
In the case of only one Higgs superfield $H_2$, anomaly cancellation
requires the presence of an extra fermion doublet
that could marry the Higgsino $\widetilde H_2$
and give a mass of order 
\begin{equation}
m_{\widetilde H}\sim \epsilon_{H_2} \ms\, .
\end{equation}
In the case of two MSSM Higgs superfields $H_{1,2}$, the two corresponding Higgsinos can marry each other
giving
\begin{equation}
m_{\widetilde H}\sim \epsilon_{H_1}\epsilon_{H_2} \ms\, .
\end{equation}
Notice that this mass term breaks any $R$-symmetry
since $H_1$ and $H_2$ must have opposite $R$-charges
in order to allow the presence of a mixing term for the scalar components, $m^2_{12}H_1H_2$.
Alternatively, we could preserve the $R$-symmetry by marrying the two Higgsinos ${\widetilde H}_{1,2}$ 
with two extra composite fermions and obtain masses of order $\epsilon_{H_{1,2}} \ms$.

Finally, in these scenarios the lightest supersymmetric particle is the gravitino (with the exception
of scenarios in which the bino gets its mass from gravitational interactions \cite{Davies:2011mp}), with a mass given by 
\be
m_{3/2}\sim \frac{F}{M_P}\sim \frac{\sqrt{N}}{4\pi}\frac{\ms^2}{M_P}
\simeq 10^{-4}\ {\rm eV}\sqrt{\frac{N}{6}} \left( \frac{\ms}{\rm 1\ TeV}\right)^2\, .
\label{gravitino}
\ee
The gravitino contains the Goldstino, which arises from the strong sector and couples to the MSSM fields with a coupling suppressed by $1/F$. If the full model has an exact $R$-parity, 
then the Goldstino is a stable particle.

\section{Matter sector}
\subsection{Elementary matter}

If the matter sector, quarks and leptons, are elementary fields not directly coupled to the 
strong sector, then the corresponding squarks and sleptons obtain their masses from 
GMSB contributions. As for gauginos, these contributions, of order ${\widetilde m}_i^2\sim{ N\ms^2}/{(16\pi^2)^2}$, are too small if $\ms\sim$ TeV.

We find only two ways to generate sizeable squark and slepton masses.
The first option is to introduce an extra $U(1)_X$ to mediate supersymmetry-breaking from the strong sector
to the MSSM. 
This extra gauge symmetry must be broken around the scale $\ms$.
Assuming that the MSSM matter fields and strong-sector fields are both charged under this $U(1)_X$, $D$-terms can mediate the supersymmetry-breaking from the strong sector to the MSSM \cite{Chacko:1999hg}. This leads to soft masses
\begin{equation}
{\widetilde m}_i^2\sim q_i g_X^2 \frac{ F^2}{M_X^2} \, ,
\label{dsofteq}
\end{equation}
where the $U(1)_X$ charges, $q_i$ are assumed to be positive for all MSSM fields and 
family-independent to avoid FCNC. If the $U(1)_X$ is broken by the strong sector, the gauge boson obtains a mass $M_X\simeq \epsilon_X\ms \simeq g_X\ms \sqrt{N}/(4\pi)$ that seems 
to be too small to avoid experimental constraints for $\ms\sim$ TeV. This then requires
that the $U(1)_X$ is broken at a slightly larger energy scale than $\ms$ by some field in the 
strong sector that does not simultaneously break supersymmetry. 
In AdS$_5$ models this larger scale could be $\Lambda_{IR}\sim N\ms$, which is related
to the 5D cut-off scale of the theory (see Appendix). 
With this assumption we obtain $M_X\sim \epsilon_X\Lambda_{IR}\sim g_X\ms N^{3/2}/(4\pi)\sim g_X\ms$ for $N\simeq 6$, allowing the mass to be a few TeV. 
Therefore, combining this with the NDA estimate of \eq{fterm}, the soft masses (\ref{dsofteq}) 
from the $D$-terms become of order
\begin{equation}
{\widetilde m}_i^2\sim q_i\frac{N \ms^2}{16\pi^2} \, .
\end{equation}
The soft masses can therefore be as large as the gaugino masses.

An alternative way to generate squark and slepton soft masses is via bilinear mixings with the strong sector, for example, $W_{int}=\lambda QU{\cal O}_{\lambda}$ where $Q,U$ are the quark superfields and ${\cal O}_\lambda$ is an operator of the strong sector. Soft masses will then be generated at the one-loop level giving 
\be
{\widetilde m}_i^2\sim \frac{\lambda^2}{16\pi^2}\ms^2\, .
\ee
We need $\lambda\gtrsim 1$ to generate masses above the experimental constraints for $\ms\sim$ TeV.

\subsubsection{The Yukawa structure}
The Yukawa couplings $Y_a$ ($a=u,d,e$) of the Higgs fields can arise, as in the MSSM,
from superpotential terms:
\be
\int d^2\theta\, \left(Y_u H_2 QU+Y_d H_1 QD+Y_e H_1 LE\right)\, .
\ee
We are therefore forced to have a two Higgs-doublet model.
This case does not suffer from FCNC, since
all the flavor structure can be encoded in the Yukawa couplings, $Y_a$.

For the case of a CFT ($\gamma_{H_i}$ constant) with a high cutoff $\Lambda_{\rm UV}$, 
we can obtain in these scenarios an upper bound on $\epsilon_{H_i}$. This is due to the
wave-function renormalization of $H_{1,2}$ arising from the mixing \eq{mixing}.
Indeed, from the RG equation for the Yukawas we obtain for $\gamma_{H_i}<0$:
\be
Y_{u,d}(\ms)\ \simeq \
Y_{u,d} (\Lambda_{\rm UV})\left(\frac{\ms}{\Lambda_{\rm UV}}\right)^{-\gamma_{H_{2,1}}}\, .
\label{rgy}
\ee
Therefore the larger $\epsilon_{H_i}^2$, the larger $|\gamma_{H_i}|$, and the smaller the Yukawa couplings at low energies. The requirement of generating sizeable Yukawa couplings
for the 3rd family quarks then places a bound on $\epsilon_{H_i}^2$.
Demanding $Y_a(\Lambda_{\rm UV})\lesssim 4\pi$, \eq{rgy} gives
\be 
\epsilon_{H_2}^2\lesssim \frac{\ln [Y_t(\ms)/(4\pi)]}{\ln [\ms/\Lambda_{\rm UV}]}\ ,\qquad
\epsilon_{H_1}^2\lesssim \frac{\ln [Y_b(\ms)/(4\pi)]}{\ln [\ms/\Lambda_{\rm UV}]}\, .
\ee
For $\Lambda_{\rm UV}/\ms\sim 10^{13}$, we have $ \epsilon_{H_2}\lesssim 0.3$ and 
$\epsilon_{H_1}\lesssim 0.5$. These constraints however can be relaxed either by taking a smaller UV cutoff $\Lambda_{\rm UV}$ or by having a non-constant $\gamma_{H_i}$ ({\it i.e.,} a departure 
from a CFT).

\subsection{Partially composite matter}

A second possibility for the matter sector is, as in the case of the Higgs,
to couple it linearly to the strong sector following \eq{mixing}.
This implies that matter superfields mix with resonances of the strong sector,
and then SM fermions are a mixture of elementary and composite fields {\it i.e.}, 
they are partially composite.
The mixing angle is of order $\epsilon_i$ ($i=Q,U,D,L,E$), defined in \eq{epsilon}.
In this case squarks and sleptons obtain masses of order 
\begin{equation}
{\widetilde m}_i\sim g_i \frac{\sqrt{N}}{4\pi}\ms\sim \epsilon_i \ms\, ,
\end{equation}
where now $g_i$ is the coupling of the matter superfield $\Phi_i$ to the strong sector.
For values of $g_i\sim 1$ these scalar masses are of the same order of magnitude as the gaugino masses.
To avoid FCNC the strong sector must preserve a flavor symmetry, implying that
$g_i$ are the same for all families.
Large values of $g_i$ make the squark and slepton masses heavier, but at the expense of increasing the degree of compositeness of the matter fields. This is bound by vertex corrections to the $W$ and $Z$ couplings,
${\delta g^{(i)}_{Z,W}}/{g^{(i)}_{Z,W}}$ that have been very well measured at LEP. We find that
\be
\frac{\delta g^{(i)}_{Z,W}}{g^{(i)}_{Z,W}}=c_{Z,W} \epsilon_{i}^2\, \frac{16\pi^2}{N} \frac{v^2}{\ms^2}\epsilon_{H_2}^2\, ,
\label{couplingratio}
\ee
where $c_{Z,W}$ is a coefficient of order one; its prediction from AdS/CFT is 
$c_{Z,W}\simeq 1.96$ (see Appendix). The strongest bounds are on the $W$ and $Z$ boson couplings to leptons and left-handed quarks that are measured at the few per mille level. For these fields, using the AdS/CFT prediction and the rough experimental bound $|{\delta g^{(i)}_{Z,W}}|/{g^{(i)}_{Z,W}}\lesssim 3\times 10^{-3}$, we obtain the constraint
\be
\epsilon_{Q,E,L}\lesssim 0.3\, \sqrt{\frac{N}{6}}\left(\frac{0.1}{\epsilon_{H_2}}\right)\left(\frac{\ms}{1\ \text{TeV}}\right)\, .
\ee
Note that for the right-handed quarks which have a small $Z$-boson coupling, these deviations are only measured at the 10$\%$ level and therefore do not put strong constraints on $\epsilon_{U,D}$.

\subsubsection{Yukawa structure and the one-Higgs doublet model}

As in the case of fully elementary matter, the Yukawa couplings can arise from the superpotential. However with partially composite matter Yukawa couplings can be generated from the strong sector. In this case, since supersymmetry is broken in the strong sector, only one Higgs doublet 
$H_2$ is required. In superfield notation the couplings can be written as
\be
\int d^2\theta\, Y_u H_2 QU+
\int d^4\theta\, \eta^\dagger \left(Y_d H_2^\dagger QD
+Y_e H_2^\dagger LE\right)\, ,
\ee
that can preserve an $R$-symmetry with $R=1$ for all matter fields and $R=0$ for the Higgs.

To avoid FCNC from four-fermion operators that, without any flavor symmetry, will only be suppressed by $g_i^4/\ms^2$ and will generically be too large, we must demand that all sources of flavor violation are proportional to the Yukawa couplings $Y_a$, {\it i.e.}, the 
minimal flavor violation (MFV) hypothesis.
In this case flavor symmetry breaking effects are parametrized by three spurion fields of the strong sector, ${\widetilde Y}_a$, that are related to the SM Yukawa couplings by 
\be
Y_u={\widetilde Y}_u\epsilon_{H_2}\epsilon_Q\epsilon_U\, ,
\label{rely}
\ee
and similarly for the other Yukawa couplings. We can now estimate the magnitude of the 
FCNC operators. For example, at the loop-level the following operator is generated 
(in superfield notation):
\be
\frac{ \epsilon_Q^4}{16\pi^2\ms^2} \int d^4\theta\, \left(Q {\widetilde Y}_u {\widetilde Y}_u^\dagger Q^\dagger \right)^2\, .
\label{fcnc}
\ee
Using the flavor constraints from Ref.~\cite{Bona:2007vi} and Eqs.~(\ref{rely}) and (\ref{fcnc}) we find that
\be
4 \pi \epsilon_{H_2}^2\epsilon^2_U \ms\gtrsim 5\ {\rm TeV}\, ,
\ee
which if satisfied implies that the up-sector must be quite composite, $\epsilon_U\sim 1$, for 
$\ms\sim $ TeV. This can have important implications in present searches
in dijet mass and angular distributions at the LHC \cite{Khachatryan:2011as}.

\section{Phenomenological implications}
\subsection{Higgs physics}

The main phenomenological implications of models with supersymmetry broken at low 
energy occur in the Higgs sector. As we have seen, these models can either have one or 
two Higgs doublets. In both cases, the lightest scalar $h$ can be much heavier than 130 GeV as shown in Figs.~\ref{mhplot} and \ref{mh2Hybrid}. It is then possible to detect this scalar 
via the golden decay channel $h \rightarrow WW/ZZ$. The strong sector also gives corrections 
to the Higgs couplings of order 
$\sim 16\pi^2 \epsilon_{H_2}^4 v^2/(N\ms^2)$ for the $hWW/ZZ$ coupling,
and $\sim 16\pi^2 \epsilon_{H_2}^2 v^2/(N\ms^2)$ for the Higgs couplings to SM fermions 
when the Yukawa terms are generated from the strong sector \cite{Giudice:2007fh}. These later
can be as large as $\sim25\%$ for $N=6$, $\ms=1$ TeV and $\epsilon_{H_2} =0.4$.

When a second Higgs doublet $H_1$ is present, as in the ordinary MSSM, there are four
extra scalars, $H, A$ and $H^\pm$. In the limit $\epsilon_{H_1}\gg \epsilon_{H_2}$
they have a common mass $m_{11}\sim\epsilon_{H_1}\ms$. The mass-splitting between 
$A$ and $H^\pm$ is given by
\be
m_{H^\pm}^2=m_A^2+\frac{2 m^2_W}{g^2}\left(2\lambda_5-\lambda_4 \right)\, ,
\ee
where we have defined $V(H_1,H_2)\supset\lambda_4 |H_1H_2|^2+\lambda_5((H_1H_2)^2+h.c.)$. These quartic couplings, which in the MSSM are given by $\lambda_4=-g_2^2/2$ and $\lambda_5=0$, receive corrections of order $\epsilon_{H_1}^2\epsilon_{H_2}^2 16\pi^2/N$, then lead to deviations from the MSSM relation:
\be
m_{H^\pm}^2-m_A^2= m_W^2+{\cal O}\left( \frac{16\pi^2}{N}\frac{v^2}{\ms^2}
\epsilon_{H_2}^2m_{H^\pm}^2\right)\, .
\ee
These deviations can be large enough to allow for the decay $H^\pm\rightarrow W^\pm A$ which could discriminate these models from ordinary MSSM models.

When the second Higgs is identified with $\Sigma$, which itself breaks electroweak symmetry, 
we find that in the limit $f_\Sigma\ll v$ only three scalars, $\pi_i$,
are lighter than $\ms$ with a mass shown in Fig.~\ref{mh2Hybrid}. If two Higgs doublets 
$H_{1,2}$ are added to $\Sigma$ we then have a three Higgs-doublet model with a potential
given in \eq{pot12s}. In the limit $f_\Sigma\ll v$ and $\epsilon_{H_{1,2}}\ll 1$ the light spectrum is similar to the MSSM (since the $\pi_i$ are heavier). However the potential \eq{pot12s} has two extra parameters $m_{i\Sigma}^2$ as compared with the MSSM. We cannot then predict the neutral Higgs masses as a function of $m_A$ and $\tan\beta$, but we can predict 
$m_{H^\pm}$ and the mixing $\sin\alpha$ which determines the mass-eigenstate Higgses $h$ and $H$ as a function of $h_1$ and $h_2$. In the first case we find the same MSSM relation 
$m_{H^\pm}^2=m_A^2+ m_W^2$, while for the second case we obtain
\be
\cos^2(\alpha-\beta)=\frac{1}{m_H^2-m_h^2}\left[m_H^2-m^2_{A}+\frac{1}{2}m^2_Z (1- 3\sin^2 2\beta)\right]\, .
\ee

\begin{figure}[top]
\centering
\includegraphics[scale=1.1]{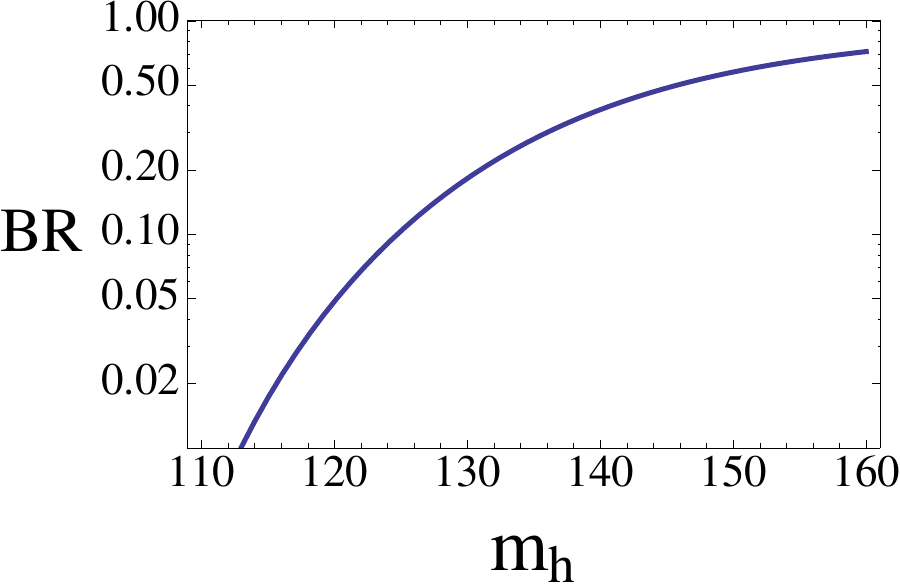} 
\caption{\footnotesize Branching ratio of $h\rightarrow \tilde G \tilde\chi_1$. See the text for details.}
\label{hgh}
\end{figure}

One of the most interesting phenomenological implications of models with low-scale supersymmetry breaking is the possible decay of the Higgs to the Goldstino and neutralino.
For the lightest Higgs this decay width is given by 
\be
\Gamma(h\rightarrow \tilde G \tilde\chi_i)= \frac{\xi_i^2}{16\pi}\frac{m_h^5}{F^2}\left[1-\left(\frac{m_{\tilde \chi_i}}{m_h}\right)^2\right]^4\, ,
\ee
where $\xi_i$ is a mixing angle \cite{Djouadi:1997gw}. This decay width is important whenever the lightest neutralino $\tilde \chi_1$ is mostly the Higgsino partner of $h$. In this case we have 
$\xi_1\simeq 1$ and BR($h\rightarrow \tilde G \tilde\chi_1$) can be sizeable if the mass of the Higgsino is around its present LEP bound $m_{\widetilde H}\gtrsim 100$ GeV. In Fig.~\ref{hgh} we show this branching ratio for $N=6$, $\ms=1$ TeV and $m_{\widetilde H}=100$ GeV.
For the other heavy Higgs see Ref.~\cite{Djouadi:1997gw}. 

In scenarios with light binos, which as explained earlier can be naturally implemented
in these models, the Higgs can also decay, via a bino-Higgsino mixing, to two binos.
Light binos, $\tilde B$, decay mostly to $\gamma+\tilde G$ but this can be outside the detector
if $m_{\tilde B}\lesssim 0.2$ GeV (for $m_{3/2}\sim 10^{-4}$ eV), or they can be even stable if $m_{\tilde B}<m_{3/2}$.
As a consequence the decay to binos corresponds to an invisible partial width for the Higgs.

\subsection{Sparticle physics and other signatures from compositeness}

If our model has a conserved $R$-parity, its phenomenology will be very similar to GMSB models. There are however a few important differences. First, the fact that the $F$-term 
can be so small,
\be
\sqrt{F}\sim 700\ {\rm GeV}
\left(\frac{N}{6}\right)^{1/4} \left(\frac{\ms}{1\ {\rm TeV}}\right)\, ,
\label{ftermv}
\ee
implies that the next-to-lightest supersymmetric particle decays inside the detector with no sizeable vertex displacement. Therefore, contrary to GMSB models, searches for superpartners must be performed in this case without requiring displaced vertices. Another important phenomenological difference is that the gauginos are Dirac states, instead of Majorana. This affects gaugino cross-sections and decays as shown in Ref.\cite{Choi:2008pi}. Nevertheless, we expect that bounds on sparticle masses from the LHC and the Tevatron on GMSB models, that are quite stringent, will be similar in our scenarios.

If $R$-parity is not conserved in our models then the bounds on the sparticles
are relaxed. The best limits to date arise from D0 where tau sneutrinos 
as light as 100 GeV are allowed~\cite{Abazov:2010km}.

A generic prediction of models with low-scale supersymmetry breaking is a light gravitino, 
(see \eq{gravitino}), that couples dominantly via the longitudinal component or Goldstino 
${\tilde G}$. If all superpartners are too heavy to be detected at the LHC, Goldstinos can still be produced with SM particles. These interactions arise from dimension-8 operators suppressed by ${1}/{F^2}$. Since ${\tilde G}$ is either stable or decays outside the detector (if $R$-parity is not conserved), the signal at hadron colliders is missing transverse momentum together with either a monojet or photon. This arises from $\{q{\bar q}, qg,gg\}\rightarrow {\tilde G}{\tilde G} +{\rm jet}$ or $q{\bar q}\rightarrow {\tilde G}{\tilde G} +\gamma$ with initial-state bremsstrahlung and thus provides a distinctive signal of our setup. The Tevatron data leads to the most stringent bound to date, $\sqrt{F} > 310$ GeV~\cite{Brignole:1998me}, while the LHC will be able to probe $\sqrt{F}$ up to approximately 2 TeV.

On the other hand, the fact that the compositeness scale $\ms\sim$ TeV is low, allows for the possible detection of resonances of the strong sector at the LHC. We expect
resonances with the same quantum numbers as the SM gauge bosons. This is similar to usual models in a warped extra dimension, except that in our scenario the fermions can all be elementary (localized on the UV boundary) and the Higgs partly composite. This implies that the resonance coupling to elementary fermions is universal and approximately given by $\sim g_i^2\sqrt{N}/(4\pi)$, where $g_i$ ($i=1,2,3$) are the SM gauge couplings, whereas the coupling to two Higgs or Goldstones behaves like $\sim \epsilon_{H_i}^2 4\pi/\sqrt{N}$. In warped spaces we can calculate these couplings and obtain $\simeq -1.15 g_i^2 \sqrt{\kappa_iN}/(4\pi) $ for fermions and $\simeq 2.25 (4\pi/\sqrt{\kappa_iN})\, \epsilon_{H_i}^2$ (in the limit $\epsilon_{H_i} \ll 1$) for 
Higgs/Goldstones.

Finally, if supersymmetry is mediated to the matter sector by a $D$-term of a $U(1)_X$ gauge group, we must have, as explained above, an extra gauge boson $Z'$ with a mass around the TeV scale. This $Z'$ boson must universally decay to leptons, which can be easily detected 
at the LHC.

\section{Conclusions}

We have shown that when supersymmetry is broken at a low scale ($\lesssim$ TeV), the Higgs 
sector of the MSSM can undergo drastic modification. The Higgs can now mix with the resonances 
of the strong sector giving rise to supersymmetry-breaking contributions to the quartic coupling of the Higgs potential. This can raise the Higgs mass not only above the tree-level MSSM value, $m_Z$ (without depending upon large radiative corrections as in ordinary MSSM models), but also for values significantly larger than $\sim 130$ GeV. In this hybrid scenario the mixing parameter interpolates between the two extremes corresponding to an elementary MSSM Higgs and a strongly-coupled Higgs resonance, with the degree of compositeness of the Higgs increasing as the Higgs becomes heavier. In this way the Higgs mass 
can range all the way up to the TeV scale, extending the usual Higgs mass range allowed in extensions of the MSSM such as NMSSM models. 

Even though the scale of supersymmetry breaking is low, a sufficiently heavy superparticle spectrum
can still be generated to avoid the ever-constraining collider bounds. For instance, gauginos
and Higgsinos can marry composite states of the strong sector, combining to form a Dirac state for which collider constraints are much less stringent. On the other hand the squarks and sleptons 
obtain masses either by mixing with composite states of the strong sector (assuming flavor symmetries to avoid FCNC) or introducing an extra U(1)$_X$ to mediate supersymmetry-breaking via flavor-independent $D$-terms.

The direct coupling of the MSSM Higgs to a strong sector generically leads to deviations
in electroweak observables which have been measured at the per-mille level. Numerical techniques at strong coupling are limited, but fortunately the AdS/CFT correspondence can be used to calculate precisely these observables in a class of strongly-coupled gauge theories with large $N$ and large t'Hooft coupling. Over most of the range of the mixing parameter the contributions to the 
$S$ and $T$ parameters are found to be within the 99$\%$ CL ellipse. The fact that a sizeable
degree of compositeness in the MSSM Higgs sector is consistent with EWPT provides an
interesting alternative to the usual MSSM.

The phenomenological implications of a heavy Higgs induced by strong dynamics include the possibility of observing the Higgs boson decay via the golden channels ($WW/ZZ$). 
In addition since a light gravitino is a generic prediction of low-scale supersymmetry breaking
a heavy Higgs can decay to a gravitino and Higgsino with a sizeable branching fraction. For 
the other MSSM Higgses the strong-coupling corrections to the MSSM relation $m_{H^\pm}^2 = m_W^2+m_A^2$ can be large enough to allow for the decay $H^\pm \rightarrow W^\pm A$ providing a distinctive signal of our scenario. Other exotic possibilities include an MSSM Higgs sector
with only one Higgs doublet if the matter sector mixes with the strong sector, obviating the 
need for a second Higgs doublet. Alternatively electroweak symmetry may actually be broken 
by the strong sector itself. In this case the VEV of $H_2$ is induced by a tadpole and large 
mixing parameters are not necessary. This possibility may occur with two Higgs doublets, effectively becoming a three Higgs doublet model. Clearly the MSSM Higgs sector can be drastically altered if it directly couples to the strong sector, providing a number of novel Higgs signatures at the LHC.

Other collider signals can arise from monojets or single photons with missing transverse
momentum associated with dimension-8 operators which couple the Goldstino to SM fields 
(assuming heavy superpartners). Furthermore the observation of resonances of the strong sector 
with quantum numbers of the SM gauge bosons, Dirac gaugino states or a TeV-scale mass $Z'$ gauge boson associated with the $U(1)_X$ gauge symmetry are also predictions of our scenario.
The possibility that both supersymmetry and supersymmetry-breaking strong dynamics are at 
the TeV scale is an intriguing option. It gives rise to a rich phenomenology that will be fully 
explored at the LHC.
$$$$

\noindent
{\bf Note added:} On December 13, 2011 the ATLAS and CMS collaborations updated
the Higgs mass exclusion limits based on the full 2011 data. In particular, the CMS collaboration now excludes a SM-like Higgs boson at the 95$\%$ CL in the mass range $127$ GeV $ \lesssim m_h \lesssim 600$ GeV. The models presented here can easily satisfy this constraint. For example, the Higgs boson can be lighter (heavier) than 127 (600) GeV if the Higgs has a small (large) amount of mixing with the strong sector. Alternatively, the Higgs bosons in our model could have escaped detection due to a reduction of the event rates. This is possible because in our model the Higgs bosons can have couplings to $WW/ZZ$ and to the top quark which are smaller than in the SM, or the Higgs boson can have an appreciable decay width into supersymmetric particles. In addition the ATLAS collaboration reported a tantalising excess near 125 GeV. If we assume that this is due to a SM-like Higgs boson, then in our setup this can be easily achieved by requiring, for example, $\tan\beta \gg 1$ with  $N=10$, $\epsilon_{H_2}=0.25$, and $c^{(22)}_\lambda=0.3$. This includes radiative corrections from a stop mass of 500 GeV that arises from choosing $\Lambda = 2~{\rm TeV}$ and $\epsilon_{Q,U}=0.25$. However there are ${\cal O}(\rm TeV)$ LHC mass limits on the gluinos and squarks that need to be avoided, and the simplest possibility to satisfy these constraints is to not assume $R$-parity.

\newpage
\section*{Acknowledgements}
The work of TG is supported by the Australian Research Council. TG thanks the IFAE, Barcelona and the SITP at Stanford for support and hospitality during the completion of this work. This material is based upon work supported in part by the National Science Foundation under Grant No.1066293 and the hospitality of the Aspen Center for Physics. The work of AP was partly supported by CICYT-FEDER-FPA2008-01430, 2009SGR894 and ICREA Academia program. AP thanks the KITP at Santa Barbara for hospitality. This research was supported in part by the National Science Foundation under Grant No. NSF PHY05-51164

\section*{Appendix: A 5D gravity description}
A quantitative analysis of the strong dynamics can be performed using the AdS/CFT 
correspondence where a strongly-coupled four-dimensional (4D) gauge theory in the large $N$ and 
large 't Hooft coupling limit is holographically related to a gravity theory in a 5D warped space.
For our purposes we will simply assume a slice of AdS$_5$ with constant radius of curvature $1/k$ 
and 5D cutoff scale $\Lambda_5$~\cite{Randall:1999ee}. The 5th dimension is labelled by $z$, with a UV (IR) brane located at the boundaries, $z=z_{UV} (z_{IR})$. The warp factor between the two boundaries, $z_{UV}/z_{IR}$ is chosen to explain the hierarchy between $\ms \sim $ TeV and the Planck scale. This corresponds to a 4D CFT with a Planck scale UV cutoff and whose conformal symmetry is spontaneously broken at $\sim \ms$. The bulk contains the SM gauge group with 5D gauge couplings $g_{{\rm 5}\, i}$ $(i=1,2,3)$. 
The 4D gauge couplings $g_i$ are related to the bulk couplings via the relation $1/g_i^2 = \log(z_{IR}/z_{UV})/(g_{{\rm 5}\, i}^2\, k) + 1/g_{IR\, i}^2+1/g_{UV\, i}^2$, where $g_{IR\, i},g_{UV\, i}$ are boundary gauge couplings. 
By comparing this equation with \eq{rung} we can identify 
\begin{equation} 
\frac{\kappa_i N}{16\pi^2} \equiv \frac{1}{g_{{\rm 5}\, i}^2\, k},
\label{Nrelation}
\end{equation}
We can choose $\kappa_2=1$ by a redefinition of $N$. In this case, with $g_2\simeq 0.65$ 
and the condition $g_{{\rm 5}\, i}^2\, k \gtrsim g_i^2 \log (z_{IR}/z_{UV})$ leads to the
bound $N\lesssim 10$. The mass of the first Kaluza-Klein (KK) gauge field can be identified with the scale $\Lambda$ and is related to the IR scale $z_{IR}^{-1}$ via 
\begin{equation}
\Lambda\equiv m_{KK}\simeq\frac{3\pi}{4} z_{IR}^{-1}\, .
\label{Lamdefn}
\end{equation}
Furthermore, to implement supersymmetry breaking at the scale $\ms$, we will assume that 
the 5D bulk and the UV-boundary are supersymmetric, with supersymmetry broken by spurion fields $\eta$ localized on the IR-boundary \cite{Gherghetta:2000kr, Hall:2003yc}.

We can use the 5D gravity description to compute the quartic coupling of an elementary Higgs field coupled to the CFT. In the limit that the 4D Higgs VEV $v\ll \Lambda$, it is a good approximation to consider the Higgs
as the zero-mode of a bulk scalar field $H$, with bulk mass $M_H$ and boundary mass 
$\Delta_H$ (in units of $k$). A massless zero mode, $h^{(0)} \sim z^{\Delta_H}$
is obtained by a supersymmetric relation, $M_H^2=\Delta_H(\Delta_H-4)$ between the bulk and boundary masses~\cite{Gherghetta:2000qt}. The dimension of the operator ${\cal O}_H$ dual to the 5D bulk Higgs is Dim$[{\cal O}_H] = 2+|\Delta_H-2|$. Using this relation, we can compute the 
dimension, $\gamma_H$ of the Higgs coupling to the strong sector. In the case of interest $-1<\gamma_H < 0$, corresponding to $2< {\rm Dim}[{\cal O}_H] < 3$ (or $-4 < M_H^2 < -3$), we can identify 
\be 
\epsilon_{H}\equiv \sqrt{-\gamma_{H}}=\sqrt{1-|\Delta_H-2|}\, .
\label{epsdefn}
\ee
Furthermore with $-4 < M_H^2 < -3$, the zero-mode behaviour ranges from $h^{(0)} \sim z$ (flat profile) to $h^{(0)} \sim z^2$ (fully composite Higgs), corresponding to
$1< \Delta_H < 2$. Our goal is to calculate the Higgs quartic coupling induced from the non-supersymmetric IR-boundary term
$V_{IR}(H)=\lambda_{IR} |H|^4$ (we are neglecting the bulk $D$-term contribution). 
This leads to the zero-mode quartic coupling
\begin{equation}
\lambda^{(0)} \simeq 16 (\Delta_H-1)^2 \lambda_{IR}k^2=16 \,\epsilon_H^4\,\lambda_{IR}k^2\, ,
\label{quartic0}
\end{equation}
where using (\ref{epsdefn}) we have $\epsilon_H^2=\Delta_H-1$.
Comparing the quartic coupling of the zero mode, $\lambda^{(0)}$ and the first excited KK state, $\lambda^{(1)}$ gives rise to the prediction
\be
\frac{\lambda^{(0)}}{\lambda^{(1)}} \simeq (\Delta_H -1)^2 \left[1-\frac{J_{1-\Delta_H}(m_1 z_{IR})J_{3-\Delta_H}(m_1 z_{IR})}{J_{2-\Delta_H}^2(m_1 z_{IR})}\right]^2\
\simeq 0.92 \,\epsilon_H^4\, ,
\label{ratioquartic}
\ee
where $m_1$ is the mass of the first scalar resonance that is approximately given by (\ref{Lamdefn}). Note also that with the above identifications the calculation of $c_{T,S}$ in (\ref{stest0}) and (\ref{stest}) using Ref.~\cite{Round:2010kj} leads to the results of \eq{csct}.

Alternatively, instead of directly computing the properties of the elementary Higgs field, we can invoke the complementary viewpoint considered in Section~\ref{alvi} and treat the Higgs field as arising from the CFT. In this way the usual AdS/CFT dictionary can be used, where the Higgs operator of the CFT is modeled by a bulk Higgs field $H$. Electroweak symmetry will then be broken by considering a generic potential on the IR boundary: $V_{IR}(H)=\lambda_{IR}(|H|^2-v_{IR}^2)^2$. The setup is similar to the gaugephobic Higgs~\cite{Cacciapaglia:2006mz}. The 5D VEV is given by
\begin{equation}
v(z) =k^{3/2} \left[\frac{2(\Delta_H-1)}{1-(z_{UV}/z_{IR})^{2(\Delta_H-1)}}
\right]^{1/2} v z_{IR}\left(\frac{z}{z_{IR}}\right)^{\Delta_H}~,
\label{v5vev}
\end{equation}
where the parameter $v_{IR}$ has been eliminated in favour of the 4D Higgs VEV $v$
and we have again neglected bulk $D$-term contributions. However to mimic the elementary Higgs we are interested in the case when $1< {\rm Dim}[{\cal O}_H] < 2$, where now 
Dim$[{\cal O}_H] = \Delta_H$. This is the alternative CFT interpretation that arises when 
$-4 < M_H^2 < -3$~\cite{Klebanov:1999tb}. The physical Higgs mass, $m_h$ and couplings can be obtained by considering fluctuations about the 5D VEV of \eq{v5vev} and deriving the effective 4D action. The quartic coupling is again found to be the same
as \eq{quartic0} except that now we have the identification 
\be
\epsilon_H^2 = {\rm Dim}[{\cal O}_H] -1~.
\ee
Similarly the zero-mode quartic coupling agrees with the result 
\eq{ratioquartic}. In the limit $m_h z_{IR} \ll 1$ the Higgs mass is found to be 
$m_h \simeq \sqrt{2\lambda^{(0)}} v$ consistent with the expression given in (\ref{mhiggs}).

Dirac gaugino masses are generated naturally in a warped extra dimension by imposing Dirichlet boundary conditions on the IR brane (assuming Neumann boundary conditions on the UV brane). Generalizing the procedure in Ref.~\cite{Gherghetta:2000kr} by 
including boundary terms we obtain
\begin{equation}
m_{\lambda_i} \simeq \frac{\sqrt{2}\, g_i}{g_{5\, i}\sqrt{k}} z_{IR}^{-1} \simeq 
g_i\frac{4\sqrt{2}}{3\pi} \frac{\sqrt{\kappa_iN}}{4\pi} \Lambda~,
\label{gauginomass}
\end{equation}
where in the second expression we have used \eq{Nrelation} and \eq{Lamdefn}. Compared to the expression \eq{mdirac}, AdS/CFT predicts $c_{\lambda_i} = \sqrt{\kappa_i} 4\sqrt{2}/(3\pi)\simeq 0.6 \sqrt{\kappa_i}$.

In the case that squark and slepton soft masses are generated from $D$-terms, 
we need an extra U(1)$_X$ in the bulk.
The simplest way to break the U(1)$_X$ symmetry is to impose Dirichlet boundary conditions on the IR brane (assuming Neumann boundary conditions on the UV brane). 
In this case the gauge boson mass is similar to that obtained for the gaugino \eq{gauginomass}. However, this value of the gauge boson mass is in conflict with experimental constraints unless $\ms\gg$ TeV. 
Alternatively we can suppose that the gauge symmetry is 
broken at the IR cutoff scale $\Lambda_{IR}=\Lambda_5 z_{UV}/z_{IR}$ so that 
the gauge boson mass $M_X \propto \Lambda_{IR}$. This 
scale can be related to $\Lambda$ by estimating when bulk perturbation theory
breaks down $\frac{g_5^2\Lambda_5}{16\pi^2} \sim 1$. This leads to the condition
\begin{equation}
\Lambda_{IR} \sim \frac{4\kappa_i}{3\pi} N \Lambda~,
\end{equation}
where we have again used \eq{Nrelation} and \eq{Lamdefn}.
Since $\Lambda_{IR}$ is larger, the experimental constraints on the $U(1)_X$ gauge boson mass can now be avoided.

If we allow fermions to propagate in the bulk then the exchange of KK states lead to vertex corrections for the gauge couplings. Modifying the result in~\cite{Agashe:2006at} 
to include an arbitrary bulk Higgs VEV, 
we obtain the corrections to the $W$ and $Z$ couplings arising from the KK gauge bosons
(neglecting hypercharge effects)
\begin{equation}
\frac{\delta g^{(i)}_{Z,W}}{g^{(i)}_{Z,W}} \simeq 9\pi^2 \frac{1/2-c_i}{128\sqrt{2}} \frac{16\pi^2}{N} \frac{v^2}{\Lambda^2} 
\frac{4(\Delta_H-1)}{\Delta_H^2}
\sim \frac{9\pi^2}{32\sqrt{2}}\, \epsilon_i^2 \frac{16\pi^2}{N} \frac{v^2}{\Lambda^2} \epsilon_H^2~,
\end{equation}
where $c_i=1/2-\epsilon_i^2$ is the 5D bulk mass of the fermion, $\epsilon_H^2 = \Delta_H-1$,
and we have assumed $c_i\ll 1/2$ and $\epsilon_H\ll 1$.
There are also corrections coming from the KK fermions that are more model dependent~\cite{Agashe:2006at}. 
Comparing to the result (\ref{couplingratio}) we find that $c_{Z,W} \simeq 1.96$.

\end{document}